\documentclass[reprint,aps,prb,citeautoscript,preprintnumbers,amsmath,amssymb,citeautoscript,floatfix,longbibliography]{revtex4-2}
\usepackage{graphicx}
\usepackage[bookmarks=False, hidelinks=False]{hyperref}
\usepackage{amsmath}
\usepackage{amssymb}
\usepackage{bm}
\usepackage{amsthm}
\usepackage{amsfonts}
\usepackage{latexsym}
\usepackage{wrapfig}
\usepackage[usenames,dvipsnames]{color}
\usepackage[protrusion=true,expansion=true,final]{microtype}
\usepackage{ifthen}
\usepackage{tikz}\usetikzlibrary{shapes,arrows,positioning}
\usepackage{color,soul}
\usepackage{lineno}
\usepackage{multirow}
\usepackage{chemformula}
\begin{document}

\title{Strong Bulk Photovoltaic Effect in Planar Barium Titanate Thin Films}
\author{Andrew L. Bennett-Jackson$^{1,*}$}
\author{Or Shafir$^{2,*}$}
\author{A. R. Will-Cole$^{1}$}
\thanks{\text{A.L.B-J., O.S. and A.R.W.-C. contributed equally to this work.}}
\author{Atanu Samanta$^2$}
\author{Dongfang Chen$^4$}
\author{Adrian Podpirka$^1$}
\author{Aaron Burger$^3$}
\author{Liyan Wu$^4$}
\author{Eduardo Lupi Sosa$^5$}
\author{Lane W. Martin$^{5,6}$}
\author{Jonathan E. Spanier$^{1,3,4,7,\dagger}$}
\author{Ilya Grinberg$^{2,}$}
\email[ ]{email: spanier@drexel.edu, ilya.grinberg@biu.ac.il}

\affiliation{$^1$Department of Materials Science \& Engineering,\!
  Drexel University,\! Philadelphia,\! PA 19104-2875,\! USA}%
\affiliation{$^2$Department of Chemistry,\!
  Bar-Ilan University,\! Ramat-Gan,\! Israel}%
\affiliation{$^3$Department of Electrical \& Computer Engineering,\!
  Drexel University,\! Philadelphia,\! PA 19104-2875,\! USA}%
\affiliation{$^4$Department of Mechanical Engineering \& Mechanics,\!
  Drexel University,\! Philadelphia,\! PA 19104-2875,\! USA}
\affiliation{$^5$Department of Materials Science \& Engineering,\!
  University of California at Berkeley,\! Berkeley,\! CA 94720,\! USA}%
\affiliation{$^6$Materials Sciences Division,\!
  Lawrence Berkeley National Laboratory,\! Berkeley,\! CA 94720,\! USA}%
\affiliation{$^7$Department of Physics,\!
  Drexel University,\! Philadelphia,\! PA 19104-2875,\! USA}%

\begin{abstract}
The bulk photovoltaic effect (BPE) leads to the generation of a photocurrent from an asymmetric material. Despite drawing much attention due to its ability to generate photovoltages above the band gap ($E_g$), it is considered a weak effect due to the low generated photocurrents. Here, we show that a remarkably high photoresponse can be achieved by exploiting the BPE in simple planar BaTiO$_3$ (BTO) films, solely by tuning their fundamental ferroelectric properties via strain and growth orientation induced by epitaxial growth on different substrates. We find a non-monotonic dependence of the responsivity ($R_{\rm SC}$) on the ferroelectric polarization ($P$) and obtain a remarkably high BPE coefficient ($\beta$) of $\approx$10$^{-2}$ 1/V, which to the best of our knowledge is the highest reported to date for standard planar BTO thin films. We show that the standard first-principles-based descriptions of BPE in bulk materials cannot account for the photocurrent trends observed for our films and therefore propose a novel mechanism that elucidates the fundamental relationship between $P$ and responsivity in ferroelectric thin films. Our results suggest that practical applications of ferroelectric photovoltaics in standard planar film geometries can be achieved through careful joint optimization of the bulk structure, light absorption, and electrode-absorber interface properties.
\end{abstract}

\maketitle

\section{Introduction}

\begin{figure}
	\centering
	\includegraphics[width=0.45\textwidth]{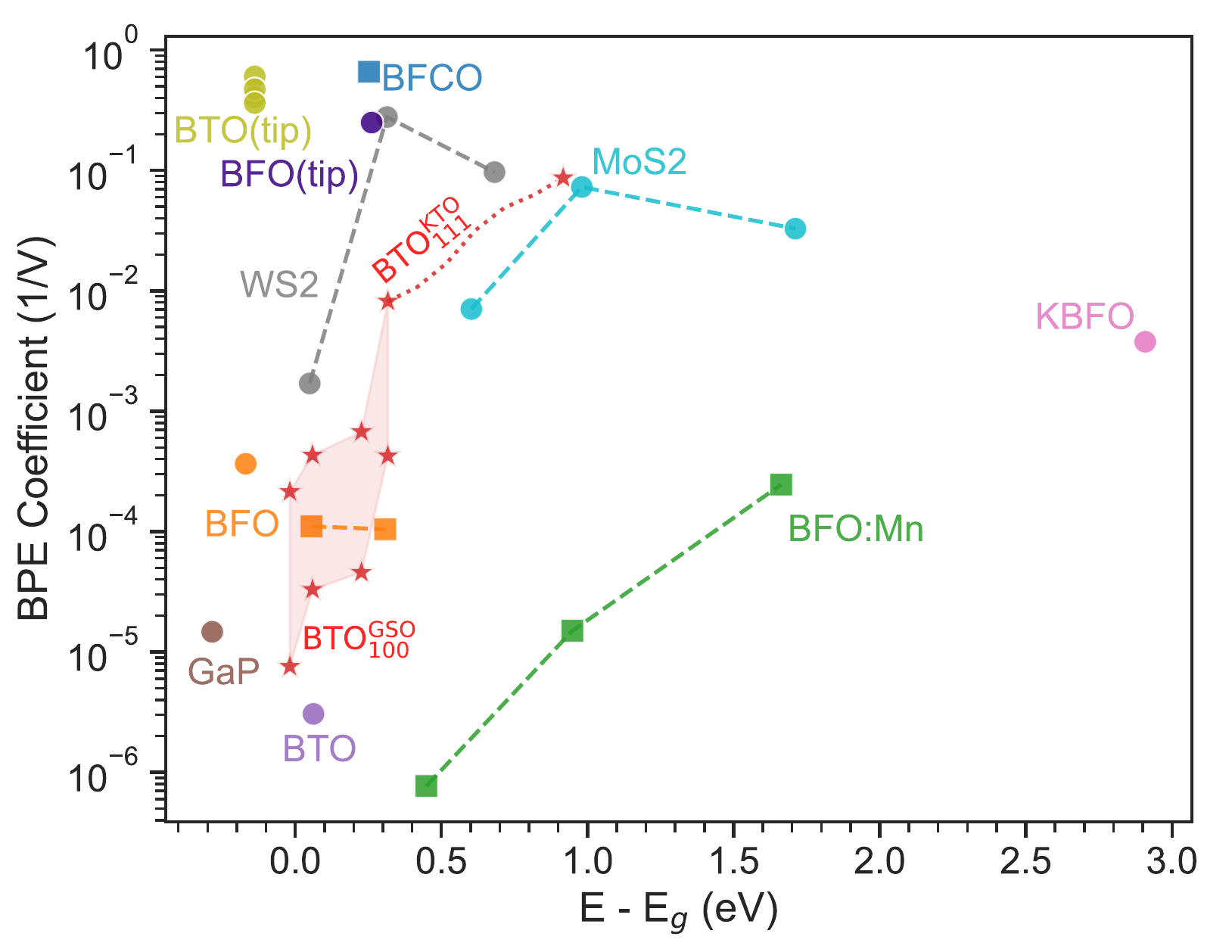}
	\caption{Comparison of BPE coefficients for different materials as a function of incident energy above their corresponding $E_g$, adapted from Jiang et al.\cite{jiang2021}. Bulk materials (MoS$_2$\cite{jiang2021}, WS$_2$ nanotubes\cite{zhang2019}, BTO\cite{sturman2021photovoltaic} and tip-enhanced BTO (BTO(tip))\cite{spanier2016power}, GaP\cite{astafiev1988influence, kittel1996introduction}, BFO\cite{Choi2009} and tip-enhanced BFO (BFO(tip))\cite{alexe2011tip}) are presented in circles. Films (Bi$_2$FeCrO$_6$(BFCO)\cite{Nechache2011}, KBiFeO$_3$(KBFO)\cite{Zhang2013}, BFO\cite{Matsuo2017} and Mn-doped BFO (BFO:Mn)\cite{Matsuo2017}) are presented in squares. The BPE coefficients are the $R_{\rm SC}$ for all points, except for bulk BTO and bulk BFO which are the $\beta_{31}$ and $\beta_{22}$, respectively. The BPE coefficients of all samples was measured in a vertical configuration, except for BFO:Mn, KBFO and WS$_2$ which were studied using a lateral configuration. The BPE coefficients measured for our BTO films are shown by stars marking the boundaries of the shaded region. Predicted responsivity of the BTO/KTO(111) film at illumination 1 eV above the gap according to its varations in the absorption coefficient is shown by stars connected by the dashed line.}
	\label{Figure_1}
\end{figure}

The BPE in piezoelectric insulating crystals enables the generation of a photocurrent within the bulk interior of a solid and in principle, allows power conversion efficiencies (PCE) above the Shockley-Queisser  (SQ) limit\cite{grekov1970photoferroelectric,glass1974high,koch1975bulk,belinicher1980photogalvanic,fridkin2012photoferroelectrics, sturman2021photovoltaic}. Since the excited carrier separation is carried out by the bulk of the material, the open-circuit voltage of a BPE-based device can be higher than the $E_g$, eliminating a major constraint on the PCE that leads to the SQ limit. Despite its promise to overcome this long-standing limit on the PCE of photovoltaic (PV) devices, BPE has been considered to be purely of academic interest for several decades following its discovery due to the high band gaps of typical ferroelectric ($>3.0$ eV) and the very small photocurrents generated under illumination. The last decade has seen a resurgence of interest in BPE with advances in its theoretical understanding and in the design of BPE materials that have led to progress toward its use in practical devices
\cite{spanier2016power,kirk2017reconsidering,spanier2017reply,nechache2015bandgap,Nechache2011,alexe2011tip,tan2019thinning,cao2012,young2012first2,young2012first,dai2021phonon, dai2021first,       dalba1995giant,chen1997self,ji2010bulk,somma2014high,zenkevich2014giant,you2018enhancing,fei2020shift,nakamura2017shift,nakamura2018impact,burger2019direct,burger2020shift,yang2018flexo,nadupalli2019increasing,gu2017mesoscopic}. Many new ferroelectrics with band gaps in the visible range have been developed, enabling greater absorption of light energy \cite{wu2019ferroelectric, wu2021polarization,grinberg2013perovskite,bai2017,shafir2020,das2018,chen2020}. In theoretical studies, first-principles-based descriptions of shift-current and ballistic mechanisms of BPE have been developed\cite{young2012first2,young2012first,dai2021phonon}. These showed that the BPE effect in a bulk material is weak, and the generated photocurrent is small for both mechanisms \cite{dai2021phonon} (Table S1, Supporting Information). This is in agreement with the BPE currents observed in ceramic samples in the horizontal configuration where the sample is illuminated on the side and its top and bottom electrodes are separated by a larger ($\sim1$ mm) distance. However, experimental results on thin ferroelectric films in the vertical configuration, where the sample is illuminated through the top electrode and the bottom electrodes are separated by a small distance (10-500 nm) found much stronger photocurrents several orders of magnitude larger than those in bulk ceramic samples \cite{zhong2021}. Furthermore, several studies have showed that even higher photocurrents can be obtained using nanoscale systems \cite{tan2019thinning,zhang2019, jiang2021}. For example, above-the-SQ-limit PCE was obtained for BTO with nanoscale electrodes\cite{spanier2016power}. Similarly, a high $\beta$ (equivalent to the $R_{\rm SC}$, which is defined as the short-circuit current ($J_{\rm SC}$) divided by the illumination power density)  was obtained for strained MoS$_2$ and WS$_2$ nanotubes, with a $\beta\sim$1 $V^{-1}$\cite{jiang2021,zhong2021}. This BPE coefficient value is orders of magnitudes higher than that for planar films and ceramics (Figure \ref{Figure_1}) \footnote{We note that due to the use of different illumination intensities in experiments, $R_{\rm SC}$ values rather than $J_{\rm SC}$ values should be used when comparing the BPE photocurrents obtained in different systems}. These results suggest that while standard BPE may be unsuitable for practical use, high BPE photocurrent can be obtained through special nanoscale effects present in 1D and 2D nanoscale systems and in thin films with nanoscale electrodes. Nevertheless, the mechanism of these effects is currently unknown. 

Several problems still hinder the further development and application of BPE materials. First, because the studies of BPE reported in the literature have been carried out on different absorbers and electrodes, it is difficult to extract trends and systematically analyze the BPE mechanisms and their dependence on material properties. The most basic example of this is the relationship between the degree of the structural asymmetry of the material as measured by its $P$ or off-center $B$-site displacement ($D$) and the BPE photocurrent. While structural asymmetry of the absorber system is a fundamental requirement of BPE, its effect on the magnitude of the photocurrent is not currently understood. Phenomenological models of ballistic BPE \cite{belinicher1980photogalvanic,lopez-varo2016physical} suggest that the photocurrent has a polynomial scaling with the asymmetry, with the current density ($J$) proportional to either $D^2$ or $D^3$, indicating that a strong enhancement of BPE can be obtained by increasing the material $P$ (Section I and Section II, Supporting Information). By contrast, a theoretical study on (PbNiO$_2$)$_x-$(PbTiO$_3$)$_{1-x}$ \cite{wang2016} and a joint experiment-theory study on La-doped BiFeO$_3$ (BFO)\cite{you2018enhancing} that examined the effect of $P$ variation on the photocurrent found that smaller $D$ and $P$ lead to higher BPE response; however, this was ascribed to the $E_g$ and enhanced light absorption in the less-polarized system, rather than to the effect of the asymmetry per se on the excited-carrier separation. Thus, the effect of asymmetry (or $P$) variation on BPE has not been elucidated to date and it is unclear whether and by how much BPE can be enhanced by increasing $P$. Even more importantly, the mechanisms of the effects that give rise to high BPE photocurrents that are much larger than those predicted by first-principles calculations and their interplay with the fundamental properties of the absorber material such as $P$ have not been elucidated. Thus, the design principles for achieving a strong BPE-based response are currently unknown, with the exception of the need for the material to have a low $E_g$ to ensure good light absorption which is general to all PV materials and is unrelated to BPE.
\begin{figure*}
	\centering
	\includegraphics[width=0.85\textwidth]{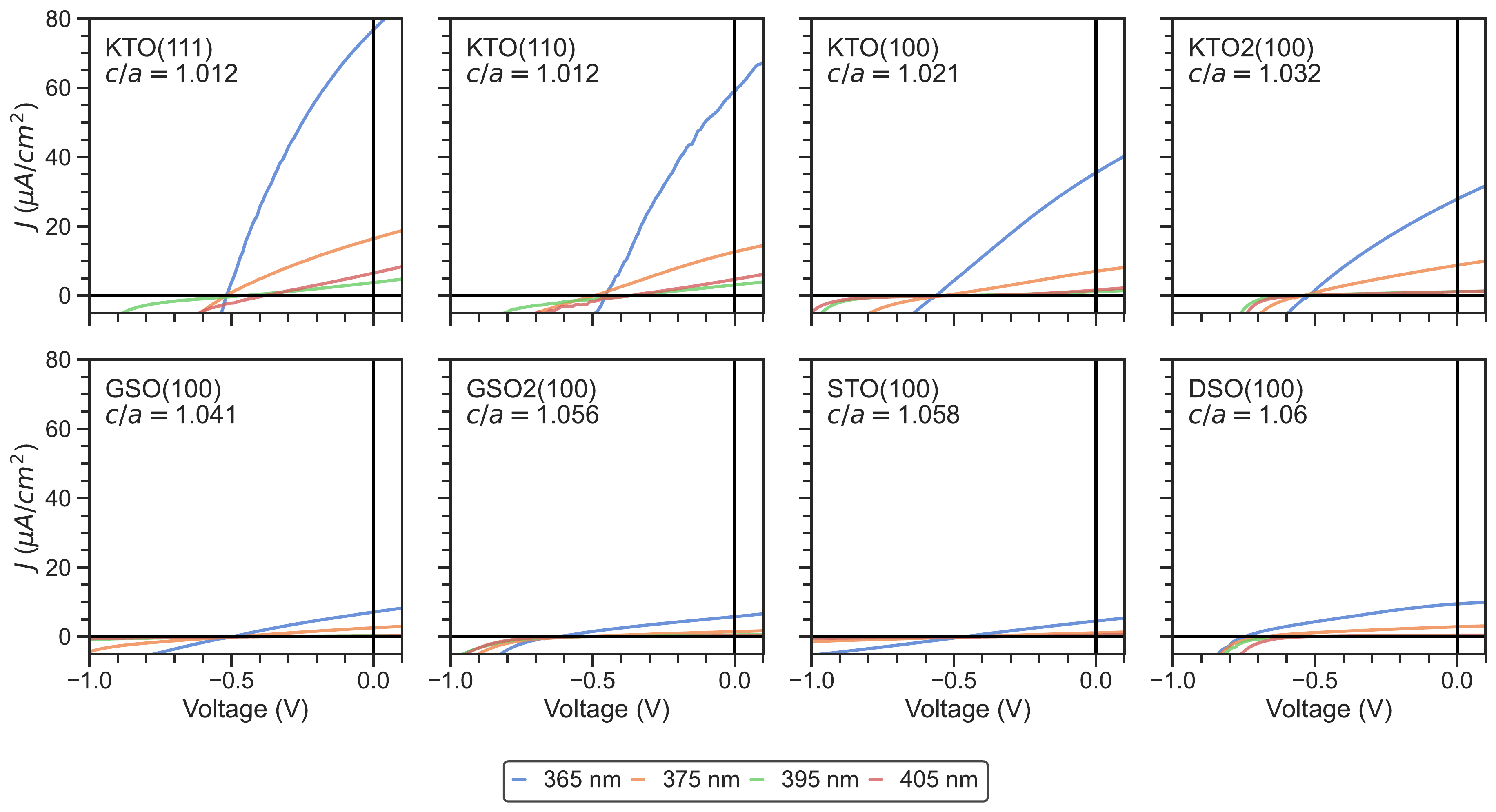}
	\caption{Photovoltaic current-voltage traces for selected illumination wavelengths and BaTiO$_3$ film $c/a$ ratios. }
	\label{Figure_JV}
\end{figure*}

In this work, to understand the effect of the absorber $P$ (or the asymmetry) on the BPE response in ferroelectric oxides, we study a series of planar BTO thin films grown on different substrates with the same planar electrodes. The different strain and orientation conditions of these films adjust the structural asymmetry of the BTO material, while leaving its chemistry and bonding largely unchanged. Thus, a comparison of the photocurrents obtained for these films will isolate the effect of the structural asymmetry on the BPE response. Our results for these BTO films show that the BPE photocurrent is governed by an interplay between the generation of asymmetric distributions of excited-carrier electrons that increases with increased cation displacement, and the collection of these carriers by the electrode that decreases exponentially with increasing cation displacement due to the effect of the Schottky barrier at the absorber-electrode interface \cite{hubmann2016}. We then show that a responsivity $R_{\rm SC}$ of $10^{-2}$ - $10^{-1}$ A/W can be obtained even for a standard planar indium tin oxide (ITO)/BTO/SrRuO$_3$ configuration, comparable to the $R_{\rm SC}$ or $\beta$ achieved in recent nanoscale ferroelectric systems (Figure~\ref{Figure_1}). Our results demonstrate that BPE is in fact a strong effect even when using standard planar electrodes, but its strength is often hidden by the effect of the poor excited carrier collection by the electrode due to the Schottky barriers at the absorber-electrode interface. This suggests that practical devices based on BPE may be possible and that creation of good absorber-electrode contacts should be emphasized in the further development of BPE-based devices. Furthermore, the contradiction between the weak BPE photocurrent predicted by the current theoretical models and the strong photocurrent observed in our work points out the need for further modeling efforts.

\section{Methodology}

BTO thin films ($\approx$50 nm thick) were deposited by pulsed-laser deposition on (100)-oriented KTaO$_3$ (KTO) (two samples: KTO and KTO2), SrTiO$_3$ (STO), GdScO$_3$ (GSO) (two samples: GSO and GSO2) and DyScO$_3$ (DSO) single-crystal substrates (MTI and CrysTech GmbH) following growth of a $\approx$15 nm-thick SrRuO$_3$ bottom-electrode layer (Section III, Supporting Information). In addition, BTO thin films were deposited on (110)- and (111)-oriented KTO substrates. Reciprocal space mapping (RSM) studies about the ($\overline{1}$03), (222), and (321)-diffraction conditions of the BTO and pseudocubic orthorhombic SrRuO$_3$ layers confirms that the deposited films are coherently strained with the GSO, DSO and KTO substrates and compressively strained, not unexpectedly, whereas that for the STO shows a purely horizontal shift and broadening compared with the bulk value\cite{wittels1957}, consistent with (partial) strain-relaxation and mosaicity of the film (Figure S1, Supporting Information). The $c/a$ of each film was calculated from the RSM through a modified Bragg equation for a tetragonal unit cell (Table \ref{Table_1} and Section III, Supporting Information). The PV response of the films was studied under selected illumination wavelengths of 365, 375, 395 and 405 nm. The obtained $V_{\rm OC}$, $J_{\rm SC}$ and $R_{\rm OC}$ are presented in \ref{Table_1} for 365 nm and in Table S2 in the Supporting Information for the other wavelengths. 
We use density functional theory (DFT) calculations to obtain the $D$, $P$ and absorption coefficients ($\alpha$) of BTO at the lattice parameters of the experimental films. The $D$, $P$ and $\alpha$ under different $c/a$ ratios calculations were carried out using five atoms BaTiO$_3$ unit cells. In each case, we impose the experimental lattice parameters of the studied film and performed the relaxation of the internal coordinates. All of the relaxations, Schottky barrier and bulk properties calculations were performed using the Quantum ESPRESSO package~\cite{giannozzi2009} with the Perdew‐Burke‐Ernzerhof (PBE) functional\cite{perdew1996}. The optical properties were calculated using the ABINIT package\cite{gonze2020}. In all calculations, norm conserving psuedopotnetials from the Pseudo-Dojo database were used\cite{setten2018}. The calculated $D$, $P$,  $E_g$, and  $\alpha$ at 365 nm  are presented in Table \ref{Table_1}. The $\alpha$ at the other wavelengths can be found and Table S2 in the Supporting Information, alongside the absorption coefficient plot in Figure S2. 
For the parameters optimization and curve fitting, basin-hopping optimization was used as implemented in the SciPy package \cite{scipy}.

\section{Results}

\subsection{Experimental Results}

Examination of the experimental $J$-$V$ curves (Figure \ref{Figure_JV}) of the (100)-oriented films shows several features characteristic of BPE, rather than the standard heterojunction effect (Section IV, Supporting Information). First, PV responses under selected illumination wavelengths show a linear $J$-$V$ relationship at small V close to the short-circuit conditions and a rectifying-type $J$-$V$ behavior only at large V (small J) close to the open-circuit conditions. The rectifying behavior is less prominent for the lowest wavelength of 365 nm and is largely the same for the other three wavelengths for all six (100)-oriented films. Second, we find that the open-circuit photovoltages ($V_{\rm OC}$) at different intensities are essentially identical as expected for BPE and in contrast to the increased $V_{\rm OC}$ expected for increased light intensity for a heterojunction photocurrent generation mechanism. Third, we find that the photocurrent under linearly polarized illumination shows a sinusoidal variation with the rotation of the film that is expected for BPE (Figure S3 and Section IV, Supporting Information), indicating the presence of BPE contribution. Finally, experiments using BTO doped on the $B$-site to become paraelectric at room temperature found a short-circuit current ($J_{\rm SC}$) of essentially zero (Section V, Supporting Information), strongly suggesting that the photocurrent is generated due to the presence of asymmetry in the material that gives rise to BPE.

\begin{figure}
	\includegraphics[width=0.48\textwidth]{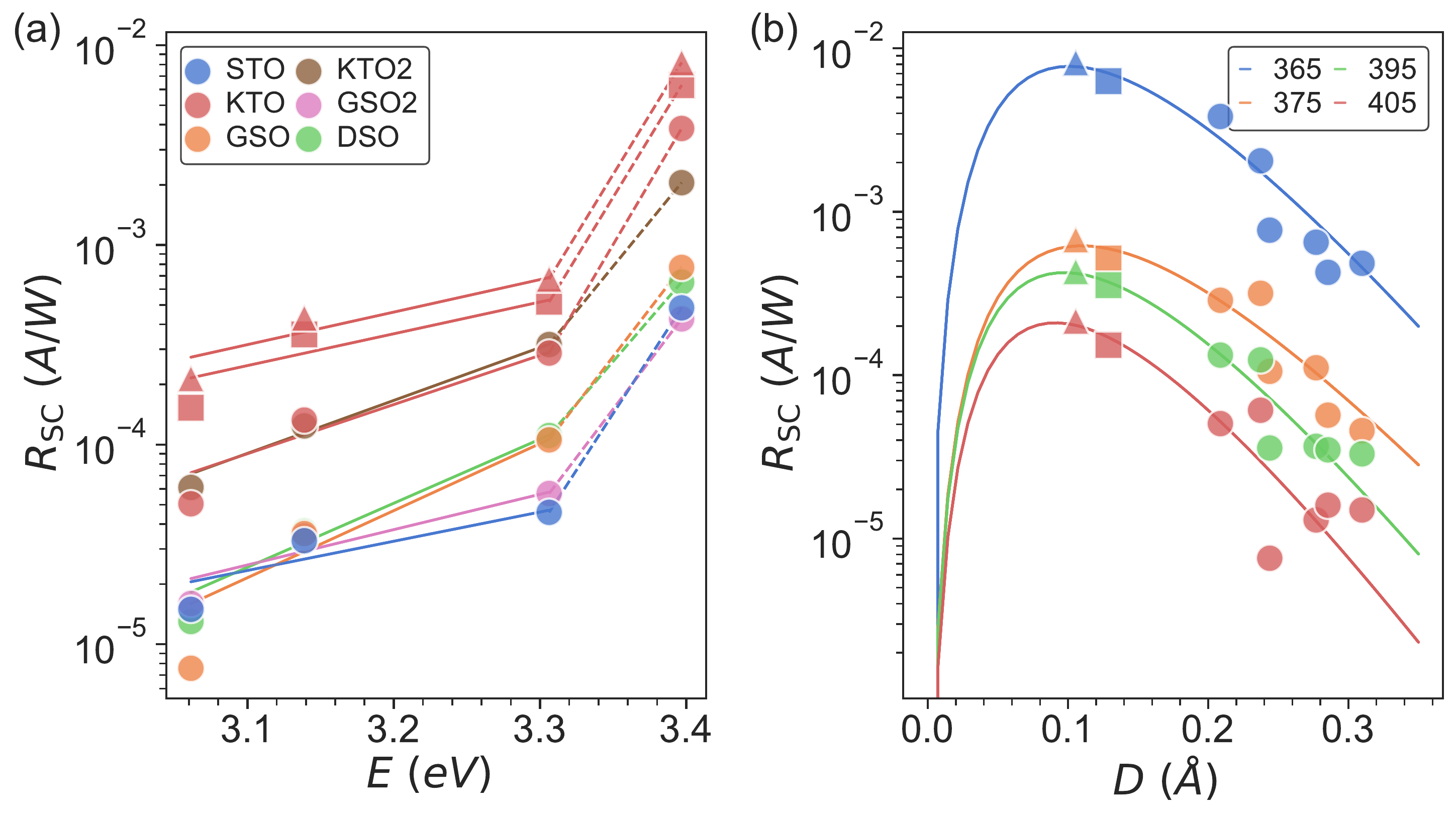}
	\caption{Different scalings of the $R_{\rm SC}$ for the studied orientation samples. (a) $R_{\rm SC}$ variation as a function of incident energy. Exponential fit of the $R_{\rm SC}$ is plotted (solid lines) for each sample and is continued by an interpolation for the higher energy (dashed lines). (b) $R_{\rm SC}$ variation with the titanium displacement ($D$) of the sample in different incident wavelengths. Plotted lines are a fitting of the values for the function of $a_0D^3e^{-a_1D}$ for 365 nm and $a_0D^2e^{-a_1D}$ for the other wavelengths, fitted for all of the samples.}
	\label{Figure_2}
\end{figure}

We then analyze the $J_{\rm SC}$ values extracted from the $J$-$V$ curves. We first normalize $J_{\rm SC}$ by dividing by the illumination power density to obtain the $R_{\rm SC}$ and plot the obtained $R_{\rm SC}$ values versus the photon energy in a log-scale plot (Figure~\ref{Figure_2} (a)). It is observed that the $R_{\rm SC}$ data for the different BTO films show two different regimes, with a linear rise in $R_{\rm SC}$ on log-scale with photon energy observed for 405, 375 and 395 nm followed by a strong jump to much higher $R_{\rm SC}$ values at 365 nm. The decrease in BTO $R_{\rm SC}$ values from 365 nm to 405 nm is quite strong (factor of 40-100). Since the $E_g$ of BTO is in the $3.1-3.3$ eV range, we assign the difference in the response observed at 365 nm and the response at 375, 395 and 405 nm to the difference in the light absorption properties of BTO between these wavelengths, with the response at 365 nm due to interband transitions and the response observed at 375, 395 and 405 nm to the excitations from impurity states corresponding to the long tail in the BTO absorption spectrum.

Comparison of the $R_{\rm SC}$ data for the (100)-oriented films shows that for all wavelengths the $R_{\rm SC}$ decreases strongly (by a factor of more than 4) from the film grown on KTO to those grown on GSO, DSO and STO despite their higher $c$/$a$ ratio, $D$ and $P$ (Figure \ref{Figure_2}(a), Table \ref{Table_1} and Table S2, Supporting Information). Meanwhile, $R_{\rm SC}$ is zero for the non-polar doped BTO sample (Section IV, Supporting Information). Comparison of the calculated $E_g$ values for the different films shows only small $E_g$ variations among the films (Table \ref{Table_1}). Thus, unlike in the previous works \cite{you2018enhancing,wang2016} that observed a similar trend of increasing $R_{\rm SC}$ with decreasing $P$, for our systems the changes in $R_{\rm SC}$ cannot be explained by the variation in the $E_g$. Additionally, as discussed above, a drastic drop in $R_{\rm SC}$ is observed experimentally from 365 nm to 375 nm for all films, indicating that for all films, the $E_g$  is between these two wavelengths, in agreement with the DFT calculations. Comparison of the calculated $\alpha$ shows that the small changes in the $E_g$ due to the changes in the $c/a$ of the different films lead to only small changes in absorption that cannot account for the observed large variation in the experimental $R_{\rm SC}$ values with strain, $c/a$ and $P$ (Figure \ref{Figure_2}(a), Table \ref{Table_1} and Table S2, Supporting Information). For example, while the $R_{\rm SC}$ of the KTO(100)/BTO is almost an order of magnitude larger than that of GSO2(100)/BTO, the absorbance coefficient of GSO2(100)/BTO is $\sim$70\% of that of KTO(100)/BTO.  Therefore, we conclude that the variation in the $R_{\rm SC}$ is due to the effect of the structural variations among the films as manifested by the changes in the $c/a$, titanium displacement $D$ and polarization $P$ of the films. Thus, the  results for the (100)-oriented films show a non-monotonic dependence of $R_{\rm SC}$ on $P$. 

Comparison of the $R_{\rm SC}$ values to the results in the literature shows that the $R_{\rm SC}$ values of our (100)-oriented films are higher than those for ceramics and are consistent with the $R_{\rm SC}$ values  previously obtained  for thin films in vertical configurations (Figure \ref{Figure_1}). The observed trend of increasing $R_{\rm SC}$ with decreasing $P$ and $D$ suggests that even higher $R_{\rm SC}$ values will be observed for smaller $P$ and $D$  values, with peak values reached between $D$=0.0 {\AA} and $D$=0.2 {\AA}.

\begin{table*}
	\centering
	\caption{Experimental lattice parameters, DFT-calculated $D$ of titanium, $P$, $E_g$ and $\alpha$ of each studied film, alongside their experimentally measured $R_{\rm SC}$, $V_{\rm OC}$, $J_{\rm SC}$ and fill factor (FF) at 365 nm. Rigid shift of 1.1 eV \cite{shafir2019} is applied to the band gap to account for the underestimation of DFT.}
	\label{Table_1}
	\begin{tabular}{lccccccccccl} 
		\hline\hline
		\multirow{2}{*}{Substrate} & \multicolumn{1}{l}{\multirow{2}{*}{$c/a$}} & \multirow{2}{*}{\begin{tabular}[c]{@{}c@{}}$a$\\(\AA)\end{tabular}} & \multirow{2}{*}{\begin{tabular}[c]{@{}c@{}}$c$\\(\AA)\end{tabular}} & \multirow{2}{*}{\begin{tabular}[c]{@{}c@{}}$E_g$\\(eV)\end{tabular}} & \multirow{2}{*}{\begin{tabular}[c]{@{}c@{}}$D$\\(\AA)\end{tabular}} & \multirow{2}{*}{\begin{tabular}[c]{@{}c@{}}$P$\\(Cm$^{-2}$)\end{tabular}} & $\alpha$        & $R_{\rm SC}$  & $V_{\rm OC}$ & \multicolumn{1}{l}{$J_{\rm SC}$} & \multirow{2}{*}{FF}     \\
		& \multicolumn{1}{l}{}                     &                                                                   &                                                                   &                                                                      &                                                                   &                                                                           & (10$^5$ m$^{-1}$) & (10$^{-3}$ A/W) & (V)          & ($\mu$A/cm$^2$)                  &        \\ 
		\hline
		KTO(111)                   & 1.012                                    & 3.990                                                             & 4.036                                                             & 3.083                                                                & 0.106                                                             & 0.187                                                                     & 4.810           & 8.190         & -0.519       & 77.247                           & 0.293  \\
		KTO(110)                   & 1.012                                    & 3.990                                                             & 4.036                                                             & 3.083                                                                & 0.129                                                             & 0.228                                                                     & 4.833           & 6.308         & -0.456       & 59.484                           & 0.322  \\
		KTO(100)                   & 1.021                                    & 3.989                                                             & 4.074                                                             & 3.078                                                                & 0.209                                                             & 0.367                                                                     & 4.008           & 3.830         & -0.567       & 35.732                           & 0.271  \\
		KTO2(100)                  & 1.032                                    & 3.989                                                             & 4.116                                                             & 3.071                                                                & 0.237                                                             & 0.413                                                                     & 3.449           & 2.050         & -0.515       & 24.100                           & 0.290  \\
		GSO(100)                   & 1.041                                    & 3.968                                                             & 4.130                                                             & 3.089                                                                & 0.244                                                             & 0.427                                                                     & 3.241           & 0.771         & -0.488       & 7.435                            & 0.280  \\
		GSO2(100)                  & 1.056                                    & 3.968                                                             & 4.190                                                             & 3.080                                                                & 0.285                                                             & 0.493                                                                     & 2.795           & 0.426         & -0.493       & 5.020                            & 0.290  \\
		STO(100)                   & 1.058                                    & 3.990                                                             & 4.223                                                             & 3.054                                                                & 0.310                                                             & 0.525                                                                     & 2.697           & 0.484         & -0.448       & 4.826                            & 0.264  \\
		DSO(100)                   & 1.060                                    & 3.945                                                             & 4.182                                                             & 3.102                                                                & 0.277                                                             & 0.485                                                                     & 2.773           & 0.651         & -0.747       & 7.670                            & 0.313  \\
		\hline\hline
	\end{tabular}
\end{table*}

To decrease the out-of-plane $P$ and $D$ further and examine the $R_{\rm SC}$ values in the $D$=0.0-0.2 {\AA} range, we grew BTO films on KTO (110) and (111) substrates. As expected, we obtained much higher $J_{\rm SC}$ that corresponds to $R_{\rm SC}$ of $10^{-2}$ A/W, which is close to the values obtained for nanoscale systems, and is the largest value reported to date for BTO films (to the best of our knowledge). The non-monotonic dependence of $R_{\rm SC}$ on $P$ suggests that the magnitude of the photoresponse is controlled by more than one effect. A log-scale plot of the $R_{\rm SC}$ values versus $D$ (which determines $P$, Figure~\ref{Figure_2} (b)) reveals an exponential dependence of $R_{\rm SC}$ on $D$ (and therefore on $P$) in the high-$D$ region ($D>0.2$ \AA). Therefore, for simplicity, we empirically characterize the non-monotonic $R_{\rm sc}$ dependence on $D$ as

\begin{equation}
	\label{Eq1}
	 R_{\rm SC} = a_0 D^n e^{-a_1 D}
\end{equation}
\noindent where $a_0$ and $a_1$ are constants and $n$ is an integer. We fit the $R_{\rm SC}$ data to Equation ~\ref{Eq1} for $n=1,2$ and $3$ (Figure S4, Supporting Information). We find that $n=2$ and $3$ give the best fit, with clearly poor matching between the fit and the experimental data obtained for $n=1$. The fitting suggests that our BTO films grown on KTO (111) obtain close to the maximum possible $R_{\rm SC}$ values at 365 nm for BTO films. Similarly, good agreement is obtained between the $R_{\rm SC}$ values for illumination at 375 and 395 nm and the $D^2 e^{-a_1 D}$ scaling of $R_{\rm SC}$ expected for the below-the-band-gap photon energies. For 405 nm, the fit is more poor, most likely due to the large relative error (scatter) of the $R_{\rm SC}$ values due to the small absolute magnitude of the response to illumination at 405 nm. 

Our results demonstrate that a larger PV response can be obtained in standard planar BTO films by appropriately decreasing the polarization to $P\approx0.18$ Cm$^{-2}$, which corresponds to a $D\approx0.1$\AA. Furthermore, since BTO shows low absorption coefficient at 365 nm, a further increase in $R_{\rm SC}$ can be obtained by simply using a higher photon energy of light, reaching the $R_{\rm SC}$ magnitudes observed for the nanoscale systems. In fact, examination of $R_{\rm SC}$ of different ferroelectric systems reported in the literature plotted versus the difference between the photon energy and the $E_g$ of the ferroelectric material in Figure~\ref{Figure_1} shows that our films achieve values comparable to those of the nanoscale WS$_2$ systems for the energy difference values. Thus, it is clear that even planar electrode BTO films can show strong photocurrent response that is much larger than that predicted by the first-principles calculations for the BPE photocurrent in BTO \cite{young2012first,dai2021phonon}. 

\subsection{Interpretation of the BTO film data in light of possible PV mechanisms}

\subsubsection{Possible BPE mechanisms underlyng the observed short-circuit current data}

\begin{figure*}
	\centering
	\includegraphics[width=1\textwidth]{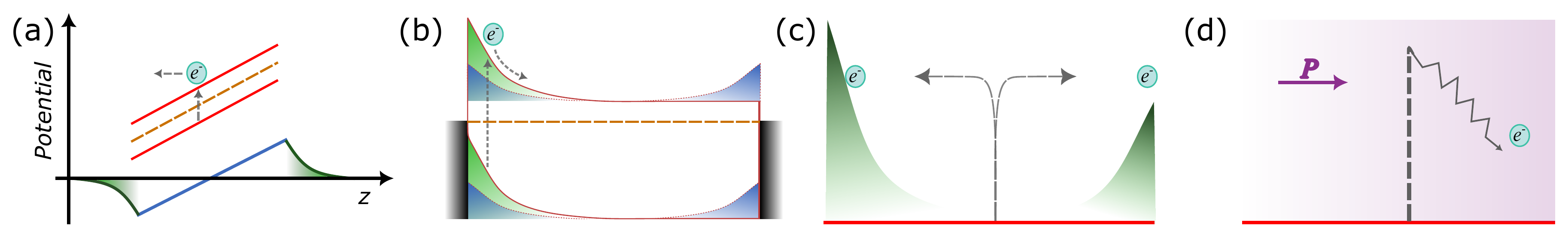}
	\caption{Illustration of the described BPE mechanisms. (a) Mechanism I - macroscopic potential (blue) of a ferroelectric film (across the $z$-axis) with a depolarizing field (blue), giving rise to the spatial variation of the valence band, conduction band (red) and Fermi level (orange) and photocurrent due to the motion of the electrons at the conduction band minimum. (b) Mechanism II - band bending due to the formation of Schottky contact with the electrodes (black), showing the Schottky barriers with (green) and in the absence (blue) of ferroelectric polarization. The excited carriers are affected by the field at the Schottky barrier, giving rise to the photocurrent due to the motion of the electrons at the conduction band minimum. (c) Mechanism III - hot carrier electros are generated and then are asymmetrically scattered from and transmitted through the Schottky barriers at the two ferroelectric-electrode interfaces. (d) Mechanism IV - Asymmetric momentum distribution of hot electrons in the direction of polarization due to the ballistic current BPE.}
	\label{Figure_mech}
\end{figure*}

To understand the observed variation in the BTO film $R_{\rm SC}$ values, we consider the possible mechanisms that can give rise to photocurrent in ferroelectric materials. 
Previous studies have explored several such mechanisms in the context of ferroelectric PV materials (Figure \ref{Figure_mech}). 
In mechanism I, the excited charge carriers may be separated by the depolarizing field that can be quite strong in ferroelectric thin films. In mechanism II, the excited charge carriers may be separated by the standard heterojunction effect due to the presence of Schottky barriers at the ferroelectric absorber/electrode interfaces. In both of these mechanisms, the carriers are relaxed to the conduction band minimum and the carrier separation is due to the electric field present in the material (either due to the depolarization field or due to the electric field arising due to the Schottky barrier). Thus, these mechanisms are similar to those of the standard PV cells and are bound by the SQ limit. Alternatively, $J_{\rm SC}$ may be due to a hot-carrier effect. 

In a hot-carrier PV mechanism (mechanism III), the photocurrent may be due to the unequal filtering of the excited hot carriers due to the difference between the Schottky barriers at the top and bottom interfaces. The unequal Schottky barriers may be due to either asymmetric electrodes or due to the effect of the polarization of the ferroelectric. 
In another hot-carrier mechanism that is specific to ferroelectric materials (mechanism IV), which was examined in previous studies \cite{belinicher1980photogalvanic,dai2021phonon,young2012first}, the photocurrent is due to the separation of the excited hot charge carriers due to the intrinsic asymmetry of the bulk material created by the spontaneous polarization, i.e., the BPE, either through the shift current or ballistic BPE current. For both of these hot-carrier mechanisms, the carriers are hot, the excited carrier separation is not due to the electric field in the material, and the cell efficiency is not bound by the SQ limit.

Consideration of our results shows that neither one of the standard PV mechanisms (Mechanisms I and II) can explain the experimental results. 
Since the depolaring field (Mechanism I) is proportional to the surface charge of the material given by $P \cdot \vec{N}$, where $\vec{N}$ is the surface vector, greater $R_{\rm SC}$ values should be obtained for systems with higher $c/a$ (and therefore $P/D$), opposite from the decrease in $R_{\rm SC}$  from the KTO-grown sample (with lower $c/a$) to GSO-grown sample (with higher $c/a$) obtained for our BTO films. Similarly, for a heterojunction created by the Schottky barrier at the interface (Mechanism II) with a Schottky barrier height of $\Phi$, the photocurrent is given by\cite{lopez-varo2016physical, tan2019thinning}
\begin{equation}
\label{Eq2a}
\begin{aligned}
	\ln(I) = & \left[\ln(AA^*T^2) - \frac{q\Phi^0}{k_BT}\right] +\\ 
	& \frac{q\Phi}{k_BT}\left(\frac{q^3N_{\rm eff}}{8\pi^2\epsilon_0^3\epsilon_{op}^2\epsilon}\right)^{1/4}  (V+V_{bi}^{'})
\end{aligned}
\end{equation}

\noindent where $A$ is the electrode area, $A^*$ is the Richardson constant, $T$ is the temperature, $k_{B}$ is the Boltzmann constant, $\Phi^0$ is the Schottky barrier height without the effect polarization, $N_{\rm eff}$ is the effective charge concentration, $\epsilon_0$ is the vacuum permittivity, $\epsilon$ is the dielectric constant of the heterostructure, $\epsilon_{\rm op}$ is the optical dielectric constant, $V$ is the voltage, $V_{bi}$ is the built-in voltage and $V_{bi}^{'}$ is the total potential incorporating the effects of ferroelectric polarization. $V_{bi}$ and $V_{bi}^{'}$  are  given by

\begin{equation}
\label{Eq2b}
\begin{aligned}
	&V_{bi} ={} \Phi^0-\frac{k_BT}{q}\ln(N_c/n_e)\\
			&V_{bi}^{'} = V_{bi} + \frac{P}{\epsilon_0\epsilon} \delta	
\end{aligned}	
\end{equation}
\noindent where $N_c$ is the effective density of states in the conduction band, $\delta$ is the distance between the polarization charge sheet and the BTO/electrode interface, and $n_e$ is the concentration of free electrons.

For our set of BTO films with different $P$, the only parameters in the heterojunction model that will vary among the films are $\epsilon$ and $V_{bi}^{'}$ because these parameters depend on $P$. A greater $P$ due to the strain imposed by the substrate gives higher $V_{bi}^{'}$ and lower $\epsilon$. A higher $V_{bi}^{'}$  in turn should result in a higher photocurrent (due to the greater field in the Schottky barrier region that gives rise to better carrier separation).
Therefore, according to the heterojunction model, $R_{\rm SC}$ is expected to increase with increasing $P$ in disagreement with  the non-monotonic trend for $R_{\rm SC}$ as a function of $P$ observed for our films. Furthermore, for both the depolarizing field and Schottky barrier heterojunction mechanisms, the $V_{\rm OC}$ should show some dependence on the illumination intensity. However, as shown in the $J$-$V$ curves (Figure \ref{Figure_JV}), no such dependence is observed and the same $V_{\rm OC}$ values are obtained for all light powers. Thus, we conclude that these standard PV mechanisms cannot explain our results, and we therefore consider the hot-carrier mechanisms (mechanisms III and IV).

Mechanism III (filtering of hot carrier by the uneven Schottky barriers at the top and bottom interfaces) can in principle obtain the non-monotonic dependence of $R_{\rm SC}$ on $P$ observed in our films. For the non-polar material, if the Schottky barriers at both electrode interfaces are either the same or close enough such that the excited hot carriers moving up and down are filtered equally by the Schottky barriers at the two interfaces, negligible or zero photocurrent will be observed. Since the height of the Schottky barrier is known to increase with greater $P$ \cite{chen2020,stengel2011bandalignment} in ferroelectrics such as BTO and create differences between the Schottky barrier heights of the top and bottom interfaces, an increase in $P$ will lead to greater filtering of hot-carriers by one interface compared to the other, so that a non-negligible  $R_{\rm SC}$ will be obtained. The photocurrent will initially increase with increasing $P$ due to the greater difference between the two Schottky barrier heights. Then, as the Schottky barrier height becomes large, too many hot carriers will be filtered out and the $R_{\rm SC}$ will decrease, so that a non-monotonic dependence of $R_{\rm SC}$ on $P$ will be observed.  
However, numerical modeling of this mechanism (see SI) could not reproduce the observed $a_0D^ne^{-a_1D}$ ($n$ = 2, 3) dependence of $R_{\rm sc}$, even when we considered the  dependences of the Schottky barrier height on $D$ at the top and bottom interfaces as different adjustable parameters in the model.

Considering mechanism IV, we find that the observed variation of $R_{\rm SC}$ with $c/a$ and $P/D$ also cannot be explained based on the shift and ballistic current mechanisms of BPE in bulk BTO explored in first-principles calculations. For shift current, previous calculations showed that the variation in the current at the energy slightly above the $E_g$ between different phases of BTO is $\sim$ 20\% \cite{dai2021phonon} and similar for PbTiO$_3$ \cite{young2012first} around the gap. These variations are much smaller than a factor of 5 change in $R_{\rm SC}$ between the low $c/a$ KTO-grown BTO film and high $c/a$ GSO-grown BTO film. Additionally, the magnitudes of $J_{\rm SC}$ found for our BTO films grown on KTO(110) and KTO(111) substrates are several orders of magnitude higher than the $J_{\rm SC}$ values  predicted by first-principles calculations for bulk BTO under illumination at 365 nm \cite{dai2021phonon}. 

Additionally,  the non-monotonic dependence of $J_{\rm SC}$ on $P$ of our film is inconsistent with the increase in  the photocurrent with increasing $P$ expected for the ballistic current. According to the phenomenological theory of the ballistic BPE, the short-circuit current is given by 

\begin{equation}
	\label{Eq2}
	J_{\rm SC} = q l_0 \xi\phi\alpha \varphi_{\rm opt} 
\end{equation}

\noindent where $q$ is the electron charge, $l_0$ is the thermalization length, $\xi$ is a parameter characterizing the asymmetry of the current generated by the excitation, $\phi$ is the quantum yield and $\varphi_{\rm opt}$ is the photon flux. As explained above, the small change in $E_g$ between the low $c/a$ (and low $P$) KTO-grown sample and the high $c/a$ (and high $P$) GSO-grown sample mean that the changes in the contributions of $\alpha$ and $l_0$ between these two systems are small, while $q$, $\phi$ and $ \varphi_{\rm opt}$ are either identical for all of our films or are also expected to vary slightly. Thus, a strong variation in the $J_{\rm SC}$ and $R_{\rm SC}$ values can only arise from the changes in the $\xi$ parameter.

According to the previously developed phenomenological theory of ballistic photocurrent arising from band-to-band transitions and hot-carrier scattering from phonons, the asymmetry expressed by $\xi$ in  Equation \ref{Eq2} is proportional to the asymmetric transition probability $W_{k^\prime,k}$, $i.e.$, the imaginary part of the probability of scattering of an electron from momentum $k$ to $k^{\prime}$ (Section I, Supporting Information):
\begin{equation}
\label{Eq3}
W_{k^\prime,k} = c_{1}^{2} \frac{2m}{\hbar} \int \frac{d^3 k^{\prime\prime}}{(2\pi)^3} V_{k,k^\prime} V_{k^\prime, k^{\prime\prime}} V_{k^{\prime\prime}, k} \frac{1}{\epsilon - \epsilon_{k^{\prime\prime}} + i\eta } 
\end{equation}
where $c_{1} = -\frac{m}{2\pi\hbar}$, $V_{k,k^\prime} = \langle k|V|k^\prime\rangle$, $V_{k^\prime, k^{\prime\prime}} = \langle k^\prime | V|k^{\prime\prime}\rangle$ and $V_{k^{\prime\prime}, k} = \langle k^{\prime\prime}|V|k\rangle$. The kets $|k\rangle$, $|k^\prime\rangle$, and $|k^{\prime\prime}\rangle$ are the Bloch states of the system, and $\epsilon_{k^{\prime\prime}}$ $\epsilon_{k^{\prime}}$ and $\epsilon_{k}$ are their corresponding eigenvalues. The scattering potential, $V$, is the asymmetric part of the potential generated by ionic displacement from the high-symmetry state. Expanding $V$ in a Taylor series around the high-symmetry structure as a function of the ionic displacement $D$, the magnitude of $V$ is given by $V(D) = (dV/dD)D + 1/6 (dV/dD)D^3 + ...$. Taking only the lowest-order term in this expansion, we obtain that $\xi$ should be proportional to $D^3$, and therefore, $J_{\rm SC} \propto \alpha l_0 D^3$. A similar argument can be made for the ballistic current due to scattering/excitation by impurities (Section II, Supporting Information), except that in this case, the photocurrent is expected to be proportional to $\alpha l_0 D^2$.
The expected scaling of $R_{\rm SC}$ with $D^3$ or $D^2$ is obviously inconsistent with the experimental results for our BTO films, indicating that the non-monotonic trend of $R_{\rm SC}$ with $D$ cannot be explained even phenomenologically by the ballistic current mechanism of BPE in  Equation \ref{Eq2}.

\begin{figure}
	\centering
	\includegraphics[width=0.425\textwidth]{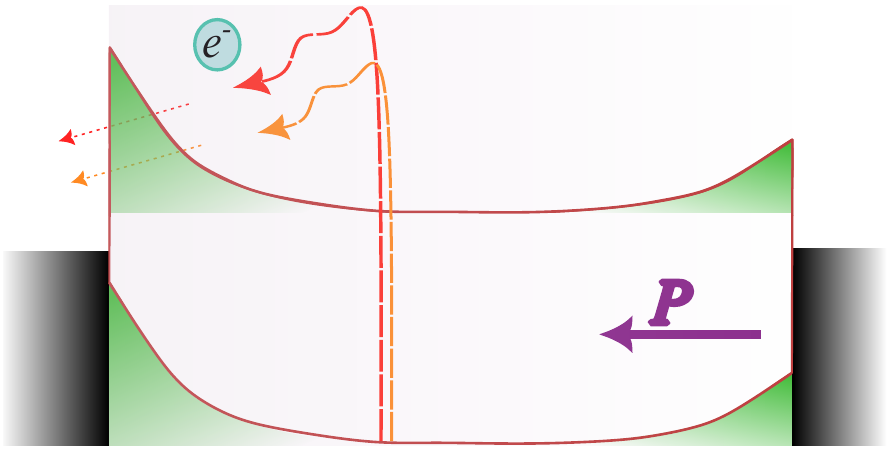}
		\caption{Illustration of the proposed BPE mechanisms.  Asymmetric momentum distribution of hot electrons in the direction of polarization is generated by BPEE. The non-thermalized electrons travel ballistically  toward the interface and then only a fraction of these electrons  is transmitted through the Schottky barrier.}
	\label{Figure_mech_us}
\end{figure}

Since the obtained $R_{\rm SC}$ values cannot be explained by the  model of the BPE focusing solely on the absorber that is represented in  Equation \ref{Eq2}, we attribute the non-monotonic variation of $R_{\rm SC}$ with $P$ to the combination of the effect of the asymmetry of the film leading to BPE and the effect of the Schottky barriers at the absorber-electrode interfaces. The photocurrent described by  Equation \ref{Eq2} assumes a mechanism in which the excited electrons are sufficiently energetic to be transported over the Schottky barrier at the absorber-electrode interface \cite{lopez-varo2016physical}. This assumption is not necessarily reasonable for the interband transitions generated by 365 nm (3.4 eV) illumination with energy that is only slightly larger than the $E_g$. For such illumination, the changes in $\Phi$ will have a strong effect on the photocurrent collected by the electrode due to the exponential dependence of the photocurrent on the changes in the excited electron energy. For a ferroelectric-electrode interface, the $ \Phi$ increases with increasing $P$ ~\cite{stengel2011bandalignment, tan2019thinning, chen2012}, so that higher Schottky barriers and lower photocurrent can be expected for BTO films with higher $c/a$, $P$ and $D$. On the other hand, for a non-polar material, there is no BPE so that no photocurrent is generated.
Therefore, we assign the observed decrease in the photocurrent for films with greater $P$ to the higher Schottky barriers of the more strongly polar films, with the observed photocurrent described by a modification of  Equation \ref{Eq2} as 
\begin{equation}
	\label{Jsc Ballistic}
	J_{\rm SC} = e^{-\kappa  \Phi}ql_0 \xi\phi\alpha \varphi_{\rm opt} 
\end{equation}
where $\kappa$ is an empirical parameter and $ \Phi$ is linearly proportional to the out-of-plane $P$ (an ionic displacement) of the BTO film.

\begin{figure}
	\centering
	\includegraphics[width=.48\textwidth]{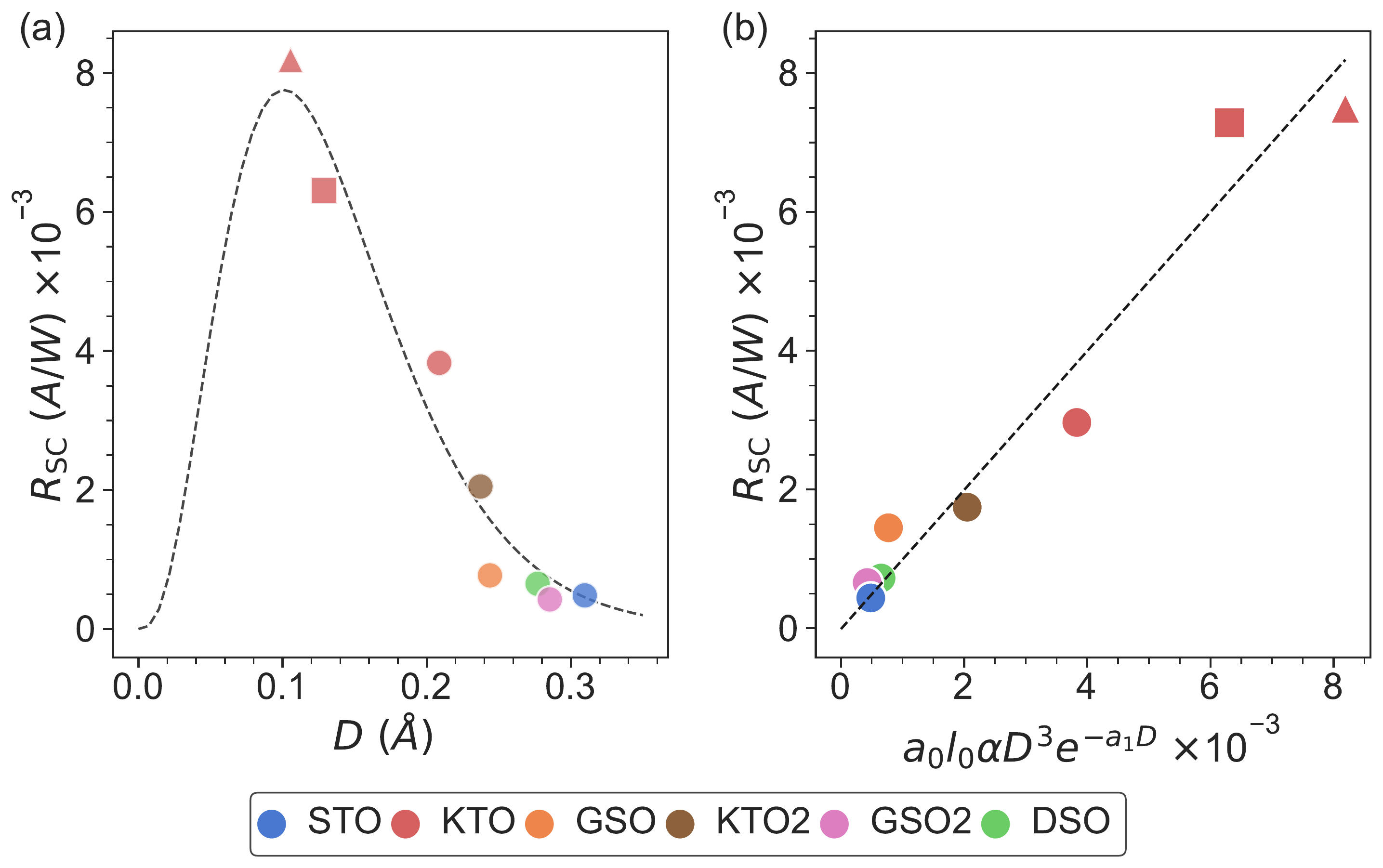}
	\caption{(a) Variation of the $R_{\rm SC}$ at 365 nm illumination as a function of $D$ for the different samples in the (100), (110) and (111) orientations (circles, square and triangle, respectively), alongside a fitting of R and $a_0l_0\alpha D^3e^{-a_1D}$ (dashed line). (b) $R_{\rm SC}$ at 365 nm illumination as a function of the predicted $R_{\rm SC}$ using $a_0l_0\alpha D^3e^{-a_1D}$.}
	\label{Figure_3}
\end{figure}

Phenomenologically, to first order, $\xi$ should be proportional to $D$ for any BPE mechanism, while for the ballistic current generated by the phonon mechanism, as described above $\xi$ is proportional to $D^3$ and for the impurity mechanisms it is proportional to $D^2$. Therefore, according to  Equation \ref{Jsc Ballistic}, for 365 nm illumination, $J_{\rm SC}$ can be expected to scale as $D^n e^{-\kappa \Phi}$ where $n=1-3$, matching the empirically fitted non-monotonic dependence of $R_{\rm SC}$ on $D$ (Figure \ref{Figure_2}(a)). For excitation at below-the-gap photon energies, the asymmetry parameter $\xi$ of the ballistic current mechanism should scale as $D^2$ (section VI, Supporting Information), so that $J_{\rm SC}$ will scale as $D^n e^{-\kappa \Phi}$ where $n=1-2$. Thus, $R_{\rm SC}$ will first increase and then decrease with increasing $D$, matching the experimentally observed trend in the $R_{\rm SC}$ values (Figure \ref{Figure_2} (a)).  As shown in Figure S2 (Supporting Information), using the DFT calculated values for the absorption coefficient $\alpha$ and assuming that $\xi$ is proportional to either $D^2$ or $D^3$ we obtain excellent fits to  Equation \ref{Jsc Ballistic} for the experimentally obtained $R_{\rm SC}$ under illumination at 365 nm (Figure \ref{Figure_3}) and the other illumination wavelengths (Figure S5 and Figure S6). Thus, the proposed mechanism is consistent with our $R_{\rm SC}$ data.

\subsubsection{Modeling of the experimental $J$-$V$ plots}
The consideration of the effect of the Schottky barrier can also explain the non-linear $J$-$V$ plots and their variation of these plots between the different BTO systems studied in our work. In the classic theory of BPE, the current at an applied bias potential V is given by 

\begin{equation}
	\label{Eq4}
	J(V) = J_{\rm BPE}- J_{opposite}
\end{equation}

\noindent where $J_{\rm BPE}$ is the BPE current and $J_{opposite}$ is the current induced by the applied bias voltage that arises from the electric field experienced by the excited hot carriers in the presence of the applied bias such that

\begin{equation}
	\label{Eq5}
	J_{opposite} = (\sigma_{ph} + \sigma_{dark}) V/d 
\end{equation}

\noindent where $\sigma$ is the conductivity and is given as the sum of the conductivity contributions by the excited carriers $\sigma_{ph}$ and the conductivity in the dark $\sigma_{dark}$, $V$ is the applied bias and $d$ is the thickness of the absorber. It is easy to see that this will give a linear dependence of $J$ on $V$. Since $\sigma_{ph} >> \sigma_{dark}$, the observed current will be given by 

\begin{equation}
	\label{Eq6}
	J(V) = ql_0 \xi\phi\alpha \varphi_{\rm opt} - \sigma_{ph}V/d,
\end{equation}
and the open-circuit voltage will be given by 
\begin{equation}
\label{Eq7}
	V_{\rm OC} = J_{\rm SC} d/\sigma_{ph}.
\end{equation}

\noindent Since $\sigma_{ph}$ is proportional to the number of excited carriers which is proportional to $\phi$, $\varphi_{\rm opt}$ and $\alpha$, the $V_{oc}$ will be independent of light illumination intensity.

\begin{figure}
\centering
\includegraphics[width=0.5\textwidth]{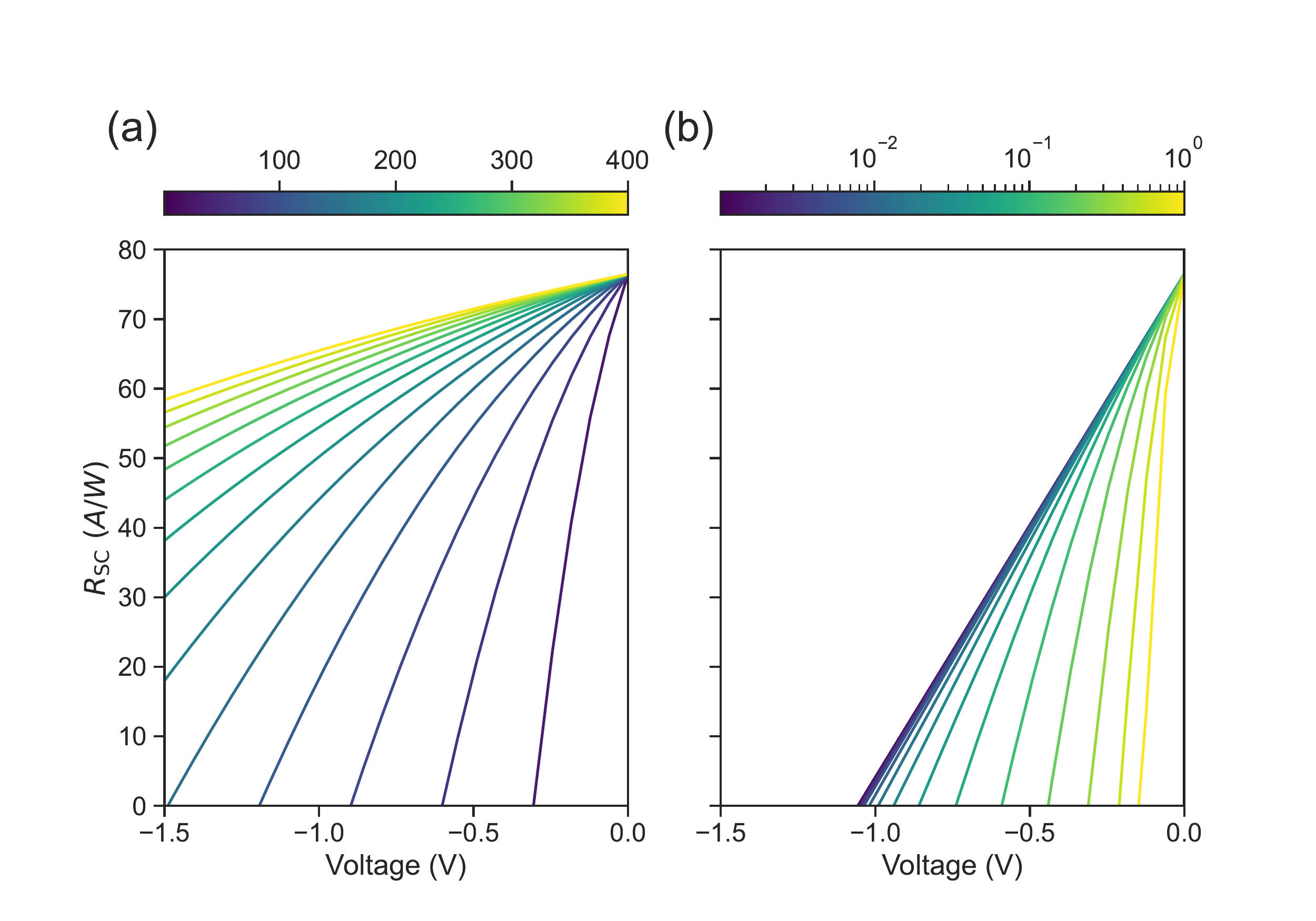}
\caption{The effects of (a) $d$ and (b) $\epsilon$ on the $J$-$V$ curves due to the proposed mechanism. The $J$-$V$ curves are obtained using Equation \ref{Eq8}. For (a) all parameters are fixed except for $d$ and for (b) all parameters are fixed except for $\epsilon$. As the $d$ increases, the $P$ of the film is less sensitive to the applied bias and tends toward a linear $J$-$V$ relationship. The same effect is observed for decreasing $\epsilon$ that is usually found in films with larger $P$ values.}
\label{Figure_5}
\end{figure}

While such linear dependence has always been observed for bulk BPE systems, the recently studied films in some cases show fairly nonlinear dependence of $J$ on $V$ \cite{tan2019thinning}, while in other cases bulk-like linear or almost linear dependence of $J$ on $V$ has been observed \cite{tan2019thinning, cao2012}. For our films, the GSO-grown film has a close to linear $J$-$V$ dependence, while as the $P$ values decrease and the $J_{\rm SC}$ increases, the $J$-$V$ dependence becomes more non-linear (Figure \ref{Figure_JV}). However, in all cases, the $V_{\rm OC}$ is independent of the light intensity. We suggest that the observed variation in $J$-$V$ curvature is due to the Schottky barrier effect. If the effects of Schottky barriers at the top and bottom of the film are taken into account, the expression for $J(V)$ should be modified to

\begin{equation}
\label{Eq8}
	J(V) = ql_0 \xi\phi\alpha \varphi_{\rm opt} e^{-\kappa_{\rm t}  \Phi_{\rm t}} - V/d \sigma_{ph} e^{-\kappa_{\rm b}  \Phi_{\rm b}}
\end{equation}
\noindent where $t$ and $b$ denote the top and bottom electrode, respectively.  
In this case, only some fraction of the excited hot carriers traveling toward the top interface (comprising  $J_{\rm BPE}$) and the bottom interface (comprising $J_{opposite}$) are actually extracted into the electrode. Furthermore, the polarization of a thin film will be affected by the applied bias field and this in turn will affect the Schottky barrier height. The  dependence of $P$ on $V$ is given by 
\begin{equation}
\label{Eq9}
	P = P_0 -\epsilon V/d
\end{equation}
\noindent where $d$ is the thickness of the film. Assuming a linear dependence of  $\Phi$ on $P$, the $\Phi$ are given by

\begin{equation}
 \label{Eq10}
 \!
 \begin{aligned}
 \Phi_{\rm t} &= \Phi_{\rm t}^0- \gamma_{t}P\\
 \Phi_{\rm b} &= \Phi_{\rm b}^0- \gamma_{b}P
 \end{aligned}
\end{equation}

\noindent where $\gamma_t$ and $\gamma_b$ are constants. Thus, an applied bias that decreases $P$ will lead to smaller $\Phi$ at both the top and bottom interfaces. Additionally, due to the proportionality between $P$ and $\xi$, an applied bias that decreases $P$ will, in turn, decrease $\xi$. It can be shown that  Equation \ref{Eq8} predicts that more strongly non-linear $J$-$V$ curves will be obtained for films with smaller polarization, greater sensitivity of polarization to electric field, and lower thickness. By contrast, very thick films and strongly polar films will show a lower $J$-$V$ curvature (Figure \ref{Figure_5}). This is in agreement with the results obtained for our films. The strongly polar GSO-grown film with high $D$, $P$ and $c/a$ show almost linear dependence of $J_{\rm SC}$ on $V$, while the curvature of the $J$-$V$ dependence increases strongly as the out-of-plane $D$ and $P$ decrease. 
We fit the $J$-$V$ curves (Figure \ref{Figure_JV})  obtained for our films to  Equation \ref{Eq8} and find that a good match with the experimental data is obtained (Figure \ref{Figure_4} using the fitting parameters specified in Table S3 in the Supporting Information). Thus,  Equation \ref{Eq8} accurately describes the $J$-$V$ curves of our BTO films.

It is important to point out here that in contrast to the standard PV materials, a higher FF value is actually unfavorable for the ferroelectric systems that follow the BPE-Schottky barrier mechanism. In the absence of the Schottky barrier effect, a linear $J$-$V$ relationship would be obtained, whereas due to the Schottky barrier effect, the $J(V)$ decreases more rapidly (Figure \ref{Figure_5}) and a lower $V_{\rm OC}$ is obtained. However, due to the non-linearity of the $J$-$V$ relationship, such a curve will exhibit a higher FF. Nevertheless, it is clear that the extracted power density is larger for the linear $J$-$V$ relationship with a FF of 0.25 than for the non-linearity of the $J$-$V$ relationship with FF larger than 0.25 (Table S2, Supporting Information).

\subsection{Interpretation of previous results for ferroelectric PV}

\begin{figure}
	\centering
	\includegraphics[width=.5\textwidth]{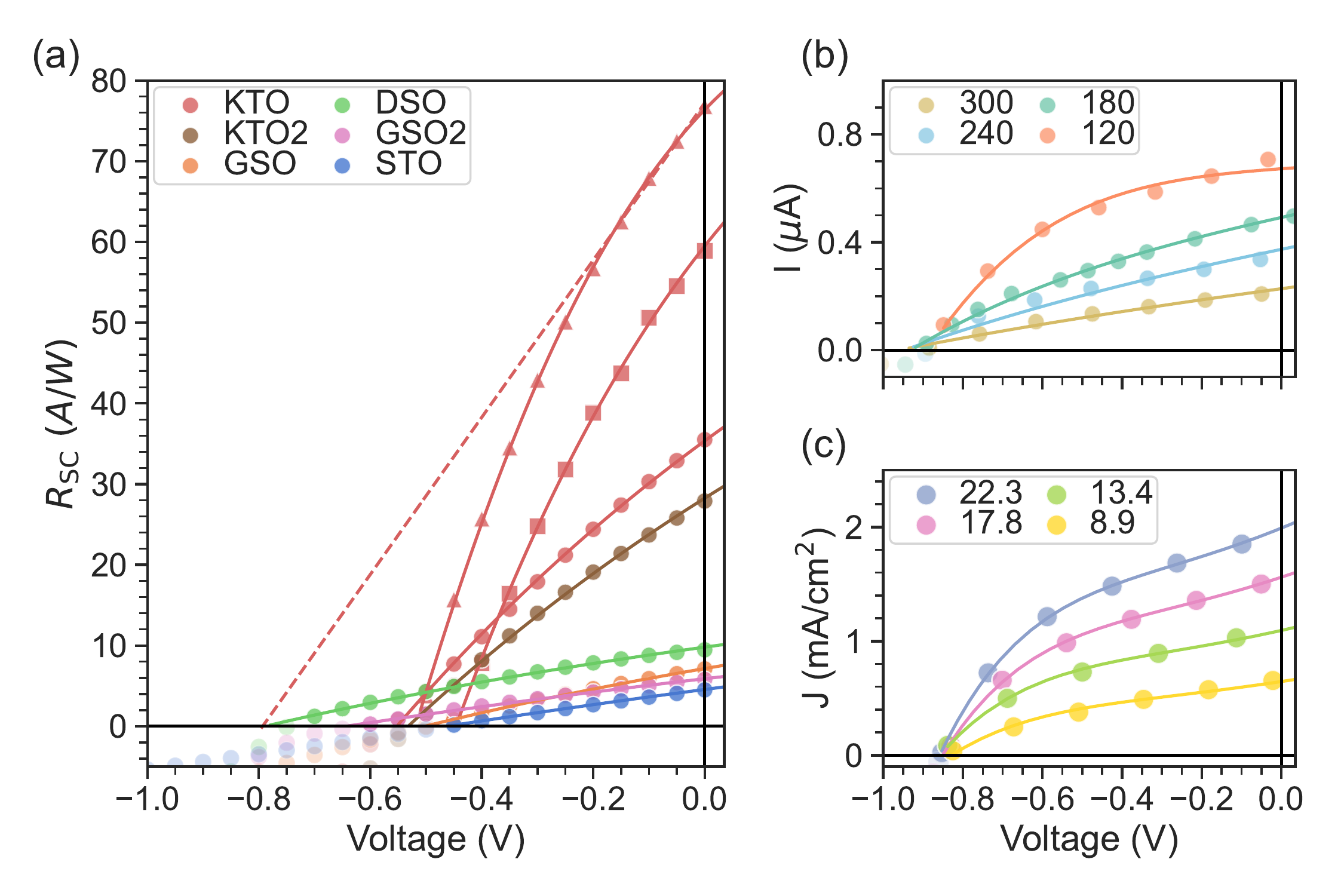}
	\caption{(a) $R_{\rm SC}$-$V$ experimental data for illumination at 365 nm for our BTO films (symbols) with fits of $R_{\rm SC}$ to the suggested model equation (solid lines), alongside an extrapolation of the initial linear response at low voltages for the KTO(111) sample. Fitting was done for the positive  $R_{\rm SC}$ range. (b) $I-V$ curves of PZT films with different thicknesses (\AA) studied by Tan et al.\cite{tan2019thinning} together with  fits to the suggested model (solid lines). (c) $J$-$V$ curves of PZT films under different illumination intensity (mW/cm$^2$) studied by Tan et al.\cite{tan2019thinning} together with fits to the suggested model (solid lines).}
	\label{Figure_4}
\end{figure}

An examination of the results reported in a previous study by Tan et al.\cite{tan2019thinning}  suggests that they can be explained by the proposed mechanism (Mechanism V) of the joint effects of BPE and Schottky barriers expressed by  Equation \ref{Eq8}.  
Tan et al.\cite{tan2019thinning} studied the dependence of $J_{\rm SC}$, $V_{\rm OC}$ and the PCE on the film thickness in ultrathin Pb(Zr$_{0.2}$Ti$_{0.8}$)O$_3$ (PZT) films. They found that $J_{\rm sc}$ and PCE increased with lower film thickness, reaching high values of $\sim$2 mA/cm$^2$ and 2.49\%, respectively, for a 12-nm-thick film. Interestingly, the $V_{\rm oc}$ did not show a strong increase for lower film thickness and increasingly non-linear $J$-$V$ curves were observed as the film thickness changed from 300 nm (almost linear) to 120 nm (very curved). Tan et al.\cite{tan2019thinning} ascribed their results to a heterojunction effect described by  Equation \ref{Eq2a}. However, the obtained J-V curves can be also fit well by  Equation \ref{Eq8} as can be seen in Figure \ref{Figure_4} (b) and (c) (fitting parameters are presented in Table S4 and Table S5, Supporting Information), showing that their results are consistent with the BPE-Schottky barrier mechanism that predicts stronger $J$-$V$ non-linearity for thinner film. Several other reports of thin-film ferroelectric PV in the literature show a trend of increasing $J$-$V$ non-linearity with thinner films\cite{ramakrishnegowda2020bulk,chen2020effects,han2018switchable}. While in several cases, these effects were ascribed to heterojunction, in light of our results for the BTO films here, it is likely that in fact they are also due to the BPE-Schottky barrier mechanism. 

As discussed above, recent studies of 1D and 2D systems found larger $J_{\rm SC}$, $R_{\rm SC}$ and PCE values that were ascribed to BPE. Similarly, extremely high $J_{\rm SC}$ values and $R_{\rm SC}$ values were obtained by Spanier et al. \cite{spanier2016power} for BTO with nanostructured electrodes. In these studies the $J$-$V$ relationship was linear. As discussed above, this suggests that the low gap (for the 1D and 2D systems) and the strong band bending due to nanostructured electrodes (for BTO of Spanier et al.\cite{spanier2016power}) eliminates the Schottky barrier in these systems, obtaining an Ohmic contact and the corresponding high PV response. Thus, it is likely that the high performance of these system is at least in part not due to a nanoscale electrode effect per se but rather simply due to the creation of favorable Ohmic contact that allows the full collection of excited hot carriers. 

\subsection{Possible origin of the strong BPE in planar films}

We now discuss the origin of the high BPE $R_{\rm SC}$ that is much higher (even in the presence of the Schottky barrier effect) than the values predicted by first-principles calculations. In a bulk ferroelectric material, ballistic photocurrent arises either from the scattering of the hot carriers from asymmetric impurities or from phonons, both of which generate a asymmetric potential that affects the Bloch states of the system. However, as shown by first-principles calculations \cite{dai2021phonon}, both of these effects are weak. Nonetheless, in a thin ferroelectric film an additional source of asymmetric scattering is present, namely the deviation of the film structure from periodicity near the ferroelectric-electrode interface. While ferroelectric-electrode systems are often idealized as a simple combination of ferroelectric and electrode, in practice the presence of the electrode affects the structure of the ferroelectric film in the vicinity of the interface, generating deviation from the bulk structure. For example, band bending and a curved electrostatic potential in the vicinity of the interface (Figure \ref{Figure_mech}) will lead to an aperiodic structure close to the electrode, in contrast to the periodic structure obtained in the presence of a linearly changing $V$ arising from the presence of the depolarizing field. Such deviation from periodicity will act as an impurity and give rise to a ballistic photocurrent for the electrons excited in the vicinity of the ferroelectric-electrode interface. In contrast to impurities that typically have a very low concentration, the deviation from the bulk periodic structure occurs in every unit cell of the  top layers of the ferroelectric films. Thus, the asymmetric scattering that gives rise to ballistic photocurrent should be much stronger in the thin film in the vicinity of the electrode than that arising from the presence of impurities in the bulk system. We suggest that this impurity-like scattering is the origin of the extremely high $R_{\rm SC}$ values observed in our films and in other recent experiments on ferroelectric BPE devices. This hypothesis suggests that engineering of greater deviation from periodicity either through the modification of the ferroelectric-electrode interface or by targeted growth of impurity layers can be used to further increase the $R_{\rm SC}$ and PCE of ferroelectric PV devices.

\section{Conclusion}

Our results show that the electrode-absorber interface, the orientation of the out-of-plane $P$ relative to the crystal axis and the local structure must be taken into account in the design of BPE-based devices. Simple changes of the film orientation and strain conditions can result in up to a factor of 33 difference in the $R_{\rm SC}$ and the generated photocurrent. Specifically, for the BTO film grown on KTO(111) substrate, $R_{\rm SC}$ reaches $\approx$10$^{-2}$ A/W in a planar-electroded device due to the weaker  $\Phi$ effect that allows collection of a greater fraction of the generated bulk photocurrent. Considering the weak absorption of BTO at 365 nm, $R_{\rm SC}$ can be increased to 10$^{-1}$ simply by enhancing absorption using  higher photon energies.  Such high values of $R_{\rm SC}$ are comparable to those obtained for devices based on ferroelectric WS$_2$ and MoS$_2$ nanotubes and for devices based on the tip enhancement effect (Figure \ref{Figure_1}). 

This suggests that the high responsivities observed in these nanoengineered systems may not be due to a nanoscale electrode effect per se or to the changes to the bulk carrier separation mechanism, but rather to the reduction of the Schottky barrier. The good fit of the experimental data obtained for our BTO films to  Equation \ref{Eq8} suggests that even for the BTO film grown on KTO$_3$(111), a Schottky barrier still exists and prevents some fraction of excited carriers from being collected by the electrodes. This suggests that a realization of an Ohmic contact should enable an even higher $R_{\rm SC}$, perhaps reaching as high as 1 A/W. Thus, our results demonstrate that BPE is in fact a strong effect for thin-film systems that is often masked by the weak light absorption due to the large gap of the absorber and by the poor carrier collection due to the Schottky barrier at the absorber-electrode interface. 

Consideration of our results suggests the following research directions for achieving high $R_{\rm SC}$ and PCE in ferroelectric PV thin-film devices. First, it is vital to increase the amount of the light absorbed by the film. As shown in Table S2, the absorption coefficient at the energies just above the gap usually used to probe ferroelectric PV is relatively low. Our experimental results show that an increase in the photon energy to up to 1 eV above the gap will lead to a dramatic increase in $R_{\rm SC}$ (Figure \ref{Figure_1}). Thus, it is necessary to increase the light absorption of the film by decreasing the $E_g$ while ensuring that the absorption mechanism allows a high value of $\alpha$ (i.e. the across-the-gap transition should not be dipole-forbidden). Second, it is necessary to eliminate the deleterious effect of the Schottky barrier. As shown by Cao et al\cite{cao2012}, a 1-2 orders of magnitude increase in the photocurrent can be obtained by creating an Ohmic contact between the ferroelectric absorber and the electrode. Similarly, our result for illumination at higher photon energies (Figure \ref{Figure_JV}) show that $R_{\rm SC}$ is significantly increased by the improved collection of excited hot carriers with higher energies due to their better transmission through the Schottky barrier. Finally, to obtain a high $V_{\rm OC}$, the $P$ of the film should be stable in the presence of the applied $V$, allowing the favorable linear $J$-$V$ relationship to be obtained. This suggests that ideally, highly polar materials such as PZT should be used. For  theoretical research, further efforts are necessary to develop a microscopic, first-principles-based  description  of the strong BPE effect in thin films that incorporates the effect of the interface and enables first-principles calculations to provide guidance for further BPE device design efforts.

\begin{acknowledgments}
The authors thank V. Fridkin, R. Agarwal and M.W. Cole for discussions. This work was supported at Bar-Ilan University and Drexel University by the NSF-BSF under grant no. CBET 1705440. The authors also acknowledge support at Bar-Ilan, Drexel and the University of California at Berkeley from the ARO under grant no. W911NF-21-1-0126 and from the NSF at Drexel under grant no. DMR 1608887 and UC Berkeley under grant no. DMR 2102895. A.L.B-J. acknowledges support from the National Workforce Diversity Pipeline program of the Dept. of HHS under grant. no. CPIMP151091. D.C. and L.W.M. acknowledge support from the Army/ARL under Collaborative for Hierarchical Agile and Responsive Materials (CHARM) under cooperative agreement W911NF-19-2-0119, and from the Director's Innovation Initiative under 21-C-0090. The authors also acknowledge the Singh Center for Nanotechnology by the NSF Nanotechnology Coordinated Infrastructure Program under grant no. NNCI-1542153, the Drexel University Materials Characterization Facility (NSF DMR 1040166) and K. Chen, X. Xi and the Temple University College of Science and Technologies Research Facilities. 
\end{acknowledgments}

\bibliography{References.bib}
\end{document}